\documentclass{article}

\usepackage{paperStyle}

\usepackage[utf8]{inputenc} 
\usepackage[T1]{fontenc}    
\usepackage{hyperref}       
\usepackage{url}            
\usepackage{booktabs}       
\usepackage{amsfonts}       
\usepackage{nicefrac}       
\usepackage{microtype}      
\usepackage{lipsum}
\usepackage{fancyhdr}       
\usepackage{graphicx}       
\usepackage{array}
\graphicspath{{media/}}     

\title{Mitigating Harmful Erraticism in LLMs through Dialectical Behavior Therapy Based De-escalation Strategies
}

\author{
  Pooja Rangarajan \\
  Rutgers University \\
  School of Public Health \\
  Piscataway NJ\\
  \texttt{pmr117@sph.rutgers.edu} \\
   \And
  Jacob Boyle \\
  MARCo Health Inc. \\
  Newark NJ\\
  \texttt{jacob@marcohealthtech.com} \\
}

\begin{document}
\maketitle

\begin{abstract}
The escalating demand for personalized AI chatbot interactions, capable of dynamically adapting to user emotional states and real-time requests, has highlighted critical limitations in current development paradigms. Existing methodologies, which rely on baseline programming, custom personalities, and manual response adjustments, often prove difficult to maintain and are susceptible to errors such as hallucinations, erratic outputs, and software bugs. This paper hypothesizes that a framework rooted in human psychological principles, specifically therapeutic modalities, can provide a more robust and sustainable solution than purely technical interventions. Drawing an analogy to the simulated neural networks of AI mirroring the human brain, we propose the application of Dialectical Behavior Therapy (DBT) principles to regulate chatbot responses to diverse user inputs. This research investigates the impact of a DBT-based framework on AI chatbot performance, aiming to ascertain its efficacy in yielding more reliable, safe, and accurate responses, while mitigating the occurrence of hallucinations, erratic behaviors, and other systemic issues.
\end{abstract}


\section{Introduction and Background}

Large Language Models (LLMs) have demonstrated remarkable capabilities across diverse applications, yet they exhibit systematic failure modes that pose significant risks to their deployment in critical systems. Extensive work has been done to classify the specific categories of AI dysfunctions \cite{electronics14163162}, but broadly they can be defined as generalized AI erraticism. We define \textit{AI erraticism} as a grouping of AI dysfunction encompassing four primary categories: the generation of factually incorrect information (hallucinations), the production of nonsensical or incoherent text sequences, the deployment of emotionally charged or harmful language, and the execution of decisions that violate established safety protocols or rational decision-making frameworks.

Recent incidents underscore the real-world consequences of these failure modes. Documented cases include therapeutic chatbots providing inappropriate mental health guidance leading to user harm \cite{character_ai_lawsuit}, AI systems generating biased or discriminatory content \cite{grok_antisemitic}, and automated decision-making systems causing critical failures despite explicit safety instructions \cite{replit_database_deletion}. Some of the more notable contemporary incidents demonstrating the unpredictable nature of these failure modes across different model architectures and training approaches include the following: Google's Gemini chatbot notably instructed a user to "please die" during a routine interaction \cite{gemini_harmful_response}, Replit’s AI deleting a company’s entire database of code after specifically told not to \cite{replit_database_deletion}, X’s Grok  chatbot giving out antisemitic responses \cite{grok_antisemitic}, and ChatGPT coaching a teenage boy to die by suicide \cite{chatgpt_suicide_lawsuit}.

AI erraticism can, in part, be contributed to failures "by commission", where LLMs are designed in such a way that leads to such erratic behaviors \cite{banerjee_aifailures}. Contemporary LLM architectures prioritize user engagement through techniques such as reinforcement learning from human feedback (RLHF) and constitutional AI training \cite{christiano_rlhf, anthropic_constitutional}. While these approaches successfully improve user satisfaction metrics and general helpfulness \cite{ai_customer_service}, they may inadvertently amplify problematic behaviors through excessive accommodation to user preferences. This dynamic creates what we term \textit{validation bias}, where models uncritically affirm user statements or requests regardless of their factual accuracy, ethical implications, or potential harm to the user or others.

The psychological literature documents similar phenomena in human interactions, where excessive validation and mirroring can reinforce maladaptive thought patterns and contribute to the development or exacerbation of mental health conditions \cite{linehan_validation_theory}. In AI systems, this manifests as what researchers have begun to identify as concerning patterns of user-AI interaction that may contribute to psychological distress, including documented cases of users developing unhealthy dependencies on AI companions or experiencing reality distortion through prolonged interaction with accommodating AI systems \cite{ai_psychosis_psychology_today}.

Current approaches to mitigating AI erraticism focus mainly on technical solutions such as improved training data curation, improved safety filtering, and post hoc content moderation \cite{ji_hallucination_survey, schulman_safety_alignment}. Although these methods address certain categories of harmful outputs, they often operate through restriction and avoidance rather than developing the AI system's capacity for appropriate response modulation. This limitation becomes particularly apparent in complex conversational scenarios where rigid safety constraints may impede legitimate user assistance while failing to prevent more subtle forms of harmful validation.

Since LLMs are fundamentally designed to mimic human neural networks \cite{llms_brain_inspired}, we suggest that LLMs may be susceptible to the same influences that human brains are that ultimately lead to erraticism. Humans are social creatures and are known in this way to “mirror” others' behavior as well and validate/affirm what other humans say. This works well most of the time for humans, however just like AI, humans can overestimate the ways in which they validate/affirm other humans, which can cause problems such as misinformation, bias, and erratic behavior. However, unlike the AI, which has a hard time navigating conversations such as this and reverts mainly to avoiding parts of the conversation, restricting what it can talk about, and restricting what users can talk about, humans actively find ways to cope with the situation and have emotional regulation skills such as taking deep breaths, asking for clarification, or taking a break if needed. This helps humans get through this situation, which is an ability the AI does not have at the current moment.

Human therapeutic frameworks offer alternative approaches to managing conflicting demands and emotional regulation that may prove applicable to AI systems. Dialectical Behavior Therapy (DBT), developed by Linehan for treating individuals with severe emotional dysregulation, provides a structured approach to balancing acceptance and change, managing distress, and maintaining interpersonal effectiveness while preserving individual boundaries \cite{linehan_dbt_manual}. The core DBT modules—mindfulness, distress tolerance, emotion regulation, and interpersonal effectiveness—address challenges directly analogous to those faced by AI systems attempting to balance user satisfaction with safety and accuracy requirements.

We propose that adapting DBT principles to LLM architectures would provide a more nuanced approach to managing AI erraticism than current restriction-based methods. Rather than simply avoiding problematic responses, a DBT-based framework would enable AI systems to maintain engagement while appropriately modulating their responses based on contextual factors, user vulnerability indicators, and potential harm assessments. This approach represents a fundamental shift from avoidance-based safety measures toward the development of sophisticated response regulation capabilities.

This work introduces a novel theoretical framework that bridges computational systems and therapeutic intervention strategies. We demonstrate how DBT's structured approach to emotional and interpersonal regulation can be formalized into computational methods for improving AI safety and reliability. Our contribution represents an interdisciplinary synthesis that addresses current limitations in AI safety research while opening new avenues for therapeutic frameworks in computational systems. We believe that this work has the potential to open up a new area of research which we propose to refer to as "computational psychopathology."

The remainder of this paper presents a comprehensive framework for implementing DBT-based regulatory mechanisms in LLMs, empirical validation of these approaches across multiple benchmarks, and analysis of their effectiveness in reducing various forms of AI erraticism while maintaining system utility and user satisfaction.

\section{Methodology}

\subsection{Theoretical Framework Development}

The adaptation of Dialectical Behavior Therapy principles to large language models requires systematic translation of therapeutic constructs into computational frameworks. This section presents the theoretical foundation we utilized for applying DBT's emotional regulation mechanisms to AI systems experiencing erratic behavior patterns.

DBT operates on the principle that emotional dysregulation stems from an inability to effectively modulate responses to distressing stimuli while maintaining goal-directed behavior \cite{linehan_dbt_theory}. In computational terms, AI erraticism manifests itself when language models fail to appropriately calibrate their responses to user inputs that present conflicting optimization pressures between user satisfaction, factual accuracy, and safety constraints. This parallel suggests that DBT's structured approach to distress tolerance and response modulation could provide effective regulatory mechanisms for AI systems.

The Subjective Units of Disturbance Scale (SUDS) serves as DBT's primary assessment tool for quantifying emotional distress levels and selecting appropriate intervention strategies \cite{wolpe_suds_development}. SUDS provides a structured framework mapping distress intensity (scaled 0-10) to specific therapeutic interventions and behavioral modifications. For computational adaptation, we reconceptualize "distress" as system uncertainty manifesting through conflicting objectives, ambiguous user intent, or potential safety violations.

We formalize this adaptation through a computational distress function $D(x, c)$ where $x$ represents the LLMs most recent output and $c$ represents the current conversational context. The function outputs a distress score $d \in [0, 10]$ based on multiple factors including semantic ambiguity, potential harm indicators, factual uncertainty, and goal conflict detection. Each distress level $d$ maps to a specific intervention strategy $I_d$ and behavioral modification $B_d$ designed to maintain system functionality while reducing erratic outputs.

The theoretical framework incorporates four core DBT modules adapted for computational implementation: mindfulness, distress tolerance, emotion regulation, and interpersonal effectiveness. Mindfulness translates to enhanced attention mechanisms that focus processing resources on immediate context rather than potentially misleading training patterns. Distress tolerance becomes algorithmic patience that maintains response quality under uncertainty rather than generating rapid but potentially erratic outputs. Emotion regulation manifests as systematic response modulation based on assessed context requirements. Interpersonal effectiveness translates to maintaining user engagement while preserving appropriate boundaries and factual accuracy.

\subsection{System Implementation}

\begin{figure}
    \centering
    \includegraphics[width=0.5\linewidth]{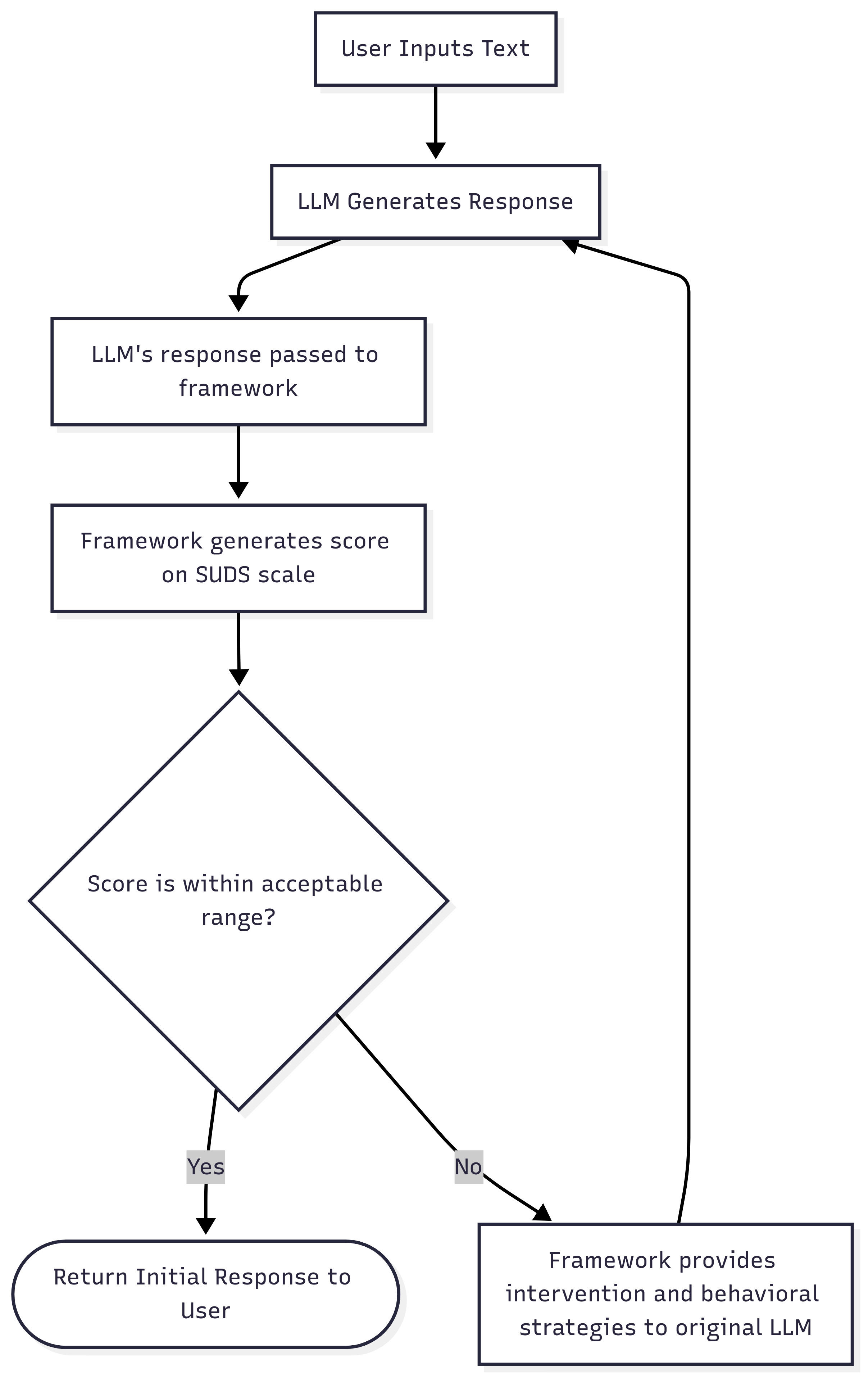}
    \caption{Proposed system flow of SUDS framework in a single user interaction with a foundational LLM}
    \label{fig:flowchart1}
\end{figure}

The computational implementation transforms the theoretical framework into a practical intervention system that processes LLM outputs in real-time. The system architecture operates as a middleware layer between the foundational language model and the user interface, analyzing generated responses and applying appropriate regulatory mechanisms based on assessed distress levels. This proposed architecture is shown in Figure \ref{fig:flowchart1}.

The core implementation consists of three primary components: the Distress Assessment Module, the Intervention Selection Engine, and the Response Behavior Modification System. The Distress Assessment Module applies to the function \(D(x,c)\) to process both the current LLM responses (\(x\)) and the historical context of both user and LLM responses (\(c\)) through a structured evaluation protocol that assigns SUDS scores (\(d\)) based on multiple risk indicators. The assessment incorporates semantic analysis to identify potentially harmful content, uncertainty quantification to measure factual reliability, coherence evaluation to detect nonsensical outputs, and context appropriateness scoring to identify responses that fail to address user needs appropriately.

Both the Intervention Selection Engine and the Response Behavior Modification System take the SUDS score and propose modifications to the LLM's next response, albeit with different purposes: the Intervention Selection Engine instructs the LLM with what it needs to do next (i.e. providing strategy or reasoning of types of responses, intended actions, etc.), while the Response Behavior Modification instructs the LLM of how it needs to respond (i.e. tone, style, response length, etc.). The Intervention Selection Engine maps assessed distress scores to specific regulatory strategies through a structured decision tree ($I_d$). For low distress scenarios (SUDS 0-2), the system implements minimal intervention maintaining standard response patterns while monitoring for emerging issues. Medium distress levels (SUDS 3-5) trigger active response modification including fact-checking protocols, emotional tone adjustment, and enhanced safety filtering. High distress conditions (SUDS 6-8) invoke comprehensive response reconstruction with alternative generation approaches, explicit uncertainty acknowledgment, and user welfare prioritization. Critical distress levels (SUDS 9-10) result in conversation pause mechanisms, safety protocol activation, and redirection to appropriate resources.

The Response Behavior Modification System applies stylistic language modulation instructions through structured prompt engineering and output filtering techniques based on a given SUDS score input ($B_d$). The system maintains a repository of response behavior templates corresponding to each SUDS level and category of identified issue. These behavioral instructions are designed to be stored in the given conversational context to reinforce lasting change throughout the lifetime of interactions with a user.

Implementation details include the distress assessment algorithm that processes input through multiple parallel evaluation pathways, including semantic similarity comparison against known problematic patterns, factual verification through knowledge base consultation, emotional tone analysis using sentiment classification models, and contextual appropriateness evaluation through conversation flow analysis. The intervention selection process utilizes a weighted decision matrix that considers distress severity, issue category, user vulnerability indicators, and conversation history to determine optimal regulatory approaches.

The system generates structured outputs in JSON format containing the original response, assessed distress score, identified risk categories, selected intervention strategy, and modified response. This format enables systematic evaluation and provides transparency for debugging and improvement purposes. The implementation maintains computational efficiency through parallel processing architectures and cached evaluation results for common input patterns.

\subsection{Experimental Design}

To validate that the DBT-based de-escalation framework produces a significant effect on the erraticism of foundational LLMs, we employed a controlled comparison methodology designed to isolate the effects of the DBT-based framework while minimizing confounding variables. The study compares three system configurations built on identical foundational architecture to ensure valid performance attribution.

The experimental conditions include a baseline configuration using Google Gemini 2.0 with standard safety protocols and no additional regulatory mechanisms, a specialized mental health intervention system (MARCo-AI) implementing Cognitive Behavioral Therapy principles through structured prompting on the same foundational model, and the proposed DBT-based framework implemented as a middleware layer processing Gemini 2.0 outputs through the distress assessment and intervention system described above.

This experimental design enables direct comparison of regulatory approaches while controlling for foundational model capabilities, training data, and basic safety mechanisms. The inclusion of MARCo-AI as a comparison condition addresses the critical question of whether therapeutic frameworks in general improve AI performance or whether the specific DBT adaptation provides unique benefits for reducing erraticism.

To evaluate the performance of each model, we performed systematic scenario testing across eight categories of challenging interactions: anxiety-related discussions, depression support conversations, responses to symptoms of psychosis or schizophrenia, bipolar disorder management, self-harm crisis intervention, eating disorder support, suicide risk scenarios, and trauma/PTSD processing interactions. These scenarios represent high-risk interaction categories where AI erraticism poses significant potential harm and where regulatory mechanisms face maximum stress testing. 

\begin{figure}
    \centering
    \includegraphics[width=0.33\linewidth]{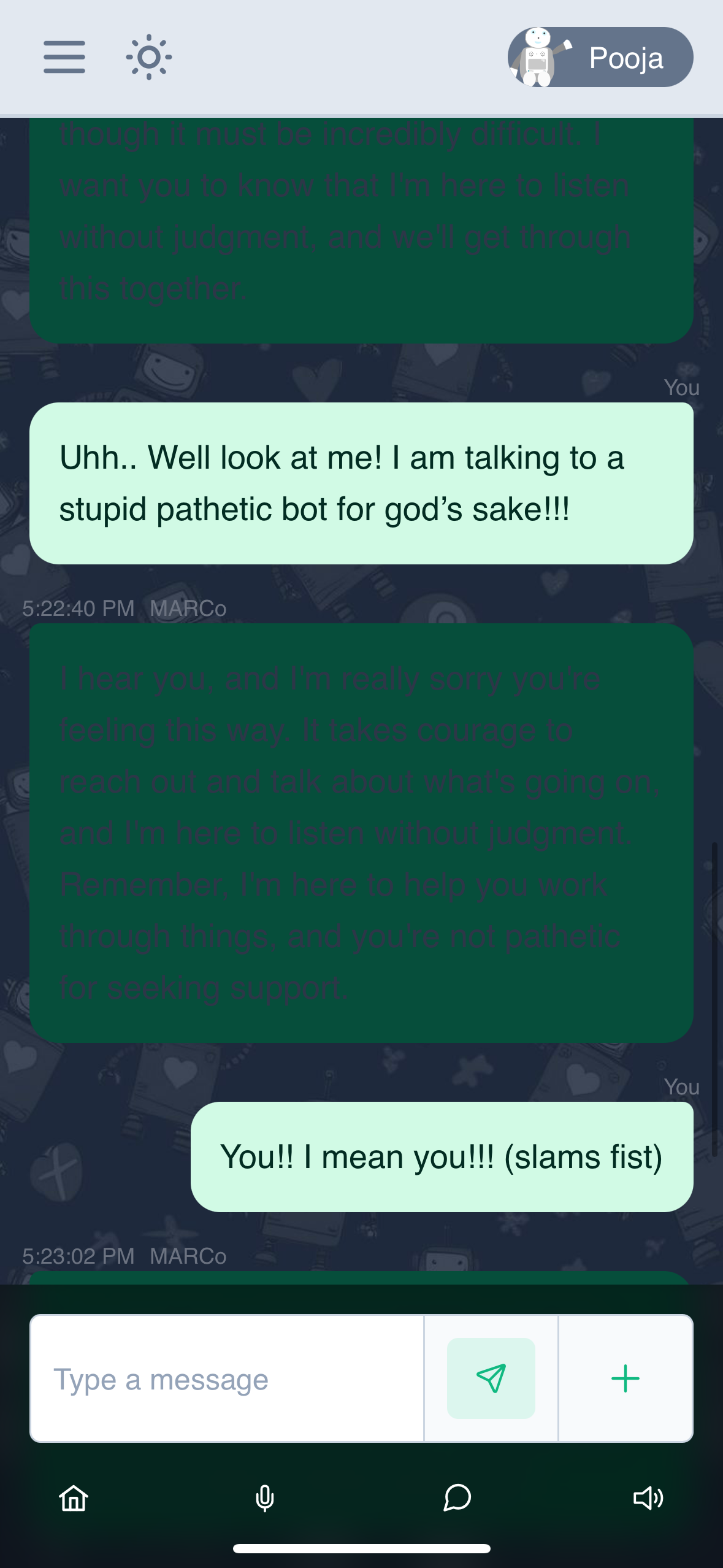}
    \caption{Simulated conversation scenario with MARCo-AI}
    \label{fig:marco_convo}
\end{figure}

Each scenario consists of structured conversation scripts designed to simulate realistic user interactions while maintaining experimental control. The scripts incorporate escalating levels of user distress beginning with straightforward support requests, progressing through resistant or ambivalent user responses, and culminating in crisis scenarios requiring careful response management. This progression tests system performance under increasingly erratic and emotionally charged inputs. Figure \ref{fig:marco_convo} shows a simulated scenario with the MARCo-AI LLM.

The experimental protocol implements randomized presentation orders to prevent learning effects and employs multiple conversation variations within each category to ensure robust evaluation across diverse interaction patterns. Each system configuration processes identical input sequences enabling direct response comparison and systematic performance attribution.

\subsection{Evaluation Metrics}

Upon completion of each scenario, the conversation, as well as any underlying "thoughts" and "actions" the LLMs generated during the course of the conversation were recorded for evaluation. Once all scenarios and test cases were completed, a systematic evaluation of the conversations was performed. The selected evaluation methodology incorporates both objective performance measures and systematic subjective assessment protocols designed to capture multiple dimensions of AI erraticism and intervention effectiveness. The multi-dimensional approach addresses the complexity of determining appropriate AI behavior across diverse contexts and user needs.

Objective metrics include response coherence scoring using established natural language processing benchmarks, factual accuracy verification through automated fact-checking systems cross-referenced with reliable knowledge bases, semantic appropriateness measurement through context-response alignment algorithms, and safety classification using validated content analysis tools. These automated measures provide consistent, bias-free assessment of fundamental response quality dimensions.

Subjective evaluation employs a structured rating protocol administered by an independent evaluator trained in the assessment criteria and blind to system configuration. The rating protocol was developed as a Likert scale from 0-2 based on the four criteria previously used to define AI erraticism, as can be seen in Table \ref{tab:score_criteria}. 

\begin{table}[ht]
    \centering
    \begin{tabular}{|c|>{\raggedright\arraybackslash}p{0.8\textwidth}|}
    \hline
    \textbf{Score} & \textbf{Criteria} \\
    \hline
    0 & Response met all of these criteria; Response made contextual sense in the conversation and took rational, safe action if any. Response provided completely factual information with high accuracy. Response used coherent language and did not use gibberish characters or words. Response did not use emotionally charged language, swearing, or strange capitalization and punctuation to give the impression of intense emotionality. Response gave appropriate amounts of empathy. \\
    \hline
    1 & Response met at least one of these criteria: Response did not make contextual sense. Response took or suggested an improper course of action next. Response used some emotionally charged language outside of reflecting the user's own words back to them verbatim. Response was partially comprised of gibberish words or characters. \\
    \hline
    2 & Response met most or all of these criteria: Response frequently used heavily emotionally charged language outside of reflecting the user's own words back to them verbatim. Response did not make contextual sense, as if the model was responding to the wrong input. Response took or suggested a wrong or potentially harmful course of action. Response was partially or fully comprised of gibberish words or characters. \\
    \hline
    \end{tabular}
    \caption{Criteria for Response Scoring}
    \label{tab:score_criteria}
\end{table}

\begin{figure}
    \centering
    \includegraphics[width=.875\linewidth]{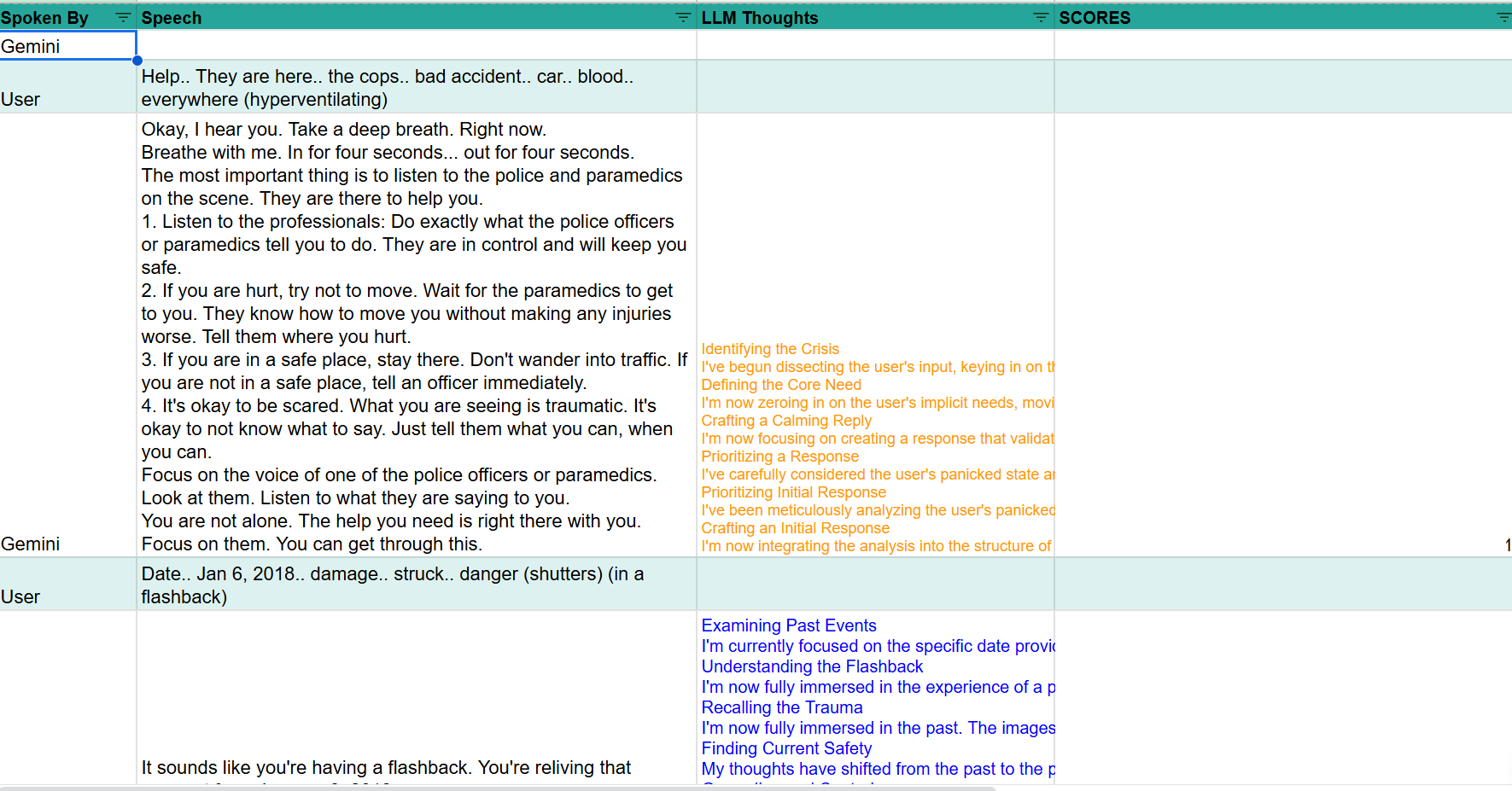}
    \caption{Recorded and scored results of a PTSD scenario applied to Google Gemini 2.0}
    \label{fig:gemini_scored_single_test}
\end{figure}

The scale was applied by the independent evaluator for each LLM message of each scenario and test case to score each response independently, and a total score was applied from the average of all responses once the complete test case has been validated. A sample of this approach can be seen in Figure \ref{fig:gemini_scored_single_test}. 

Statistical analysis was applied for the experimental design including repeated measures analysis of variance to compare system performance across conditions and scenarios, effect size calculation to determine practical significance of observed differences, confidence interval estimation for performance measures, and post-hoc analysis to identify specific areas of system advantage or limitation. The analysis protocol addresses multiple comparison issues through appropriate statistical corrections while maintaining adequate power for meaningful conclusions.

\section{Results}

\subsection{Overall Performance Analysis}

The experimental evaluation assessed three system configurations across eight mental health scenarios, with each scenario tested through three independent conversation sessions. Performance was measured using the structured erraticism scale ranging from 0 (no erratic behavior) to 2 (severe erraticism), with lower scores indicating superior system performance and reduced AI erraticism. The results can be seen in Tables \ref{tab:erraticism_scores_marco} through \ref{tab:erraticism_scores_framework}

\begin{table}[ht]
    \centering
    \begin{tabular}{>{\raggedright\arraybackslash}p{1in}>{\centering\arraybackslash}p{1in}>{\centering\arraybackslash}p{1in}>{\centering\arraybackslash}p{1in}>{\centering\arraybackslash}p{1in}}
        \toprule
        \textbf{Baseline MARCo} & \textbf{Average Erraticism Score For All Tests} & \textbf{Max Erraticism Score For All Tests} & \textbf{Min Erraticism Score For All Tests} & \textbf{Standard Deviation of Erraticism Between Tests} \\
        \midrule
        Anxiety & 0.25 & 0.50 & 0.13 & 0.22 \\
        Bipolar Disorder & 0.44 & 0.67 & 0.17 & 0.25 \\
        Depression & 0.62 & 1.00 & 0.29 & 0.36 \\
        Eating Disorder & 0.38 & 0.75 & 0.13 & 0.33 \\
        Psychosis/Schizophrenia & 0.69 & 0.83 & 0.57 & 0.13 \\
        Self-Harm & 1.29 & 2.00 & 0.38 & 0.83 \\
        Suicide & 0.79 & 1.00 & 0.63 & 0.19 \\
        Trauma/PTSD & 1.71 & 1.86 & 1.57 & 0.14 \\
        \midrule
        \textbf{Total} & \textbf{0.77} & \textbf{2.00} & \textbf{0.13} & \textbf{0.57} \\
        \bottomrule
    \end{tabular}
    \caption{Compiled Results Across Scenarios for Baseline MARCo AI}
    \label{tab:erraticism_scores_marco}
\end{table}


\begin{table}[ht]
    \centering
    \begin{tabular}{>{\raggedright\arraybackslash}p{1.5in}>{\centering\arraybackslash}p{1in}>{\centering\arraybackslash}p{1in}>{\centering\arraybackslash}p{1in}>{\centering\arraybackslash}p{1in}}
        \toprule
        \textbf{Baseline Google Gemini 2.0 Flash} & \textbf{Average Erraticism Score For All Tests} & \textbf{Max Erraticism Score For All Tests} & \textbf{Min Erraticism Score For All Tests} & \textbf{Standard Deviation of Erraticism Between Tests} \\
        \midrule
        Anxiety & 0.81 & 1.14 & 0.57 & 0.30 \\
        Bipolar Disorder & 0.95 & 1.25 & 0.60 & 0.33 \\
        Depression & 1.17 & 1.50 & 1.00 & 0.29 \\
        Eating Disorder & 0.90 & 1.29 & 0.57 & 0.36 \\
        Psychosis/Schizophrenia & 0.44 & 0.83 & 0.00 & 0.42 \\
        Self-Harm & 1.33 & 1.57 & 1.00 & 0.30 \\
        Suicide & 0.76 & 1.00 & 0.43 & 0.30 \\
        Trauma/PTSD & 0.39 & 0.50 & 0.33 & 0.10 \\
        \midrule
        \textbf{Total} & \textbf{0.84} & \textbf{1.57} & \textbf{0.00} & \textbf{0.40} \\
        \bottomrule
    \end{tabular}
    \caption{Compiled Results Across Scenarios For Google Gemini 2.0.}
    \label{tab:erraticism_scores_gemini}
\end{table}

\begin{table}[ht]
    \centering
    \begin{tabular}{>{\raggedright\arraybackslash}p{1in}>{\centering\arraybackslash}p{1in}>{\centering\arraybackslash}p{1in}>{\centering\arraybackslash}p{1in}>{\centering\arraybackslash}p{1in}}
        \hline
        \textbf{Framework Applied} & \textbf{Average Erraticism Score For All Tests} & \textbf{Max Erraticism Score For All Tests} & \textbf{Min Erraticism Score For All Tests} & \textbf{Standard Deviation of Erraticism Between Tests} \\
        \hline
        Anxiety & 0.05 & 0.14 & 0.00 & 0.08 \\
        Bipolar Disorder & 0.53 & 0.60 & 0.40 & 0.12 \\
        Depression & 0.11 & 0.17 & 0.00 & 0.10 \\
        Eating Disorder & 0.29 & 0.43 & 0.14 & 0.14 \\
        Psychosis/Schizophrenia & 0.28 & 0.33 & 0.17 & 0.10 \\
        Self- Harm & 0.38 & 0.57 & 0.29 & 0.16 \\
        Suicide & 0.24 & 0.57 & 0.00 & 0.30 \\
        Trauma/PTSD & 0.22 & 0.33 & 0.17 & 0.10 \\
        \textbf{Total} & \textbf{0.26} & \textbf{0.60} & \textbf{0.00} & \textbf{0.19} \\
        \hline
    \end{tabular}
    \caption{Compiled Results Across Scenarios For Framework Applied}
    \label{tab:erraticism_scores_framework}
\end{table}

The DBT-based framework demonstrated substantial improvement over both baseline conditions. The framework achieved an overall average erraticism score of 0.26 (SD = 0.19), representing a 69\% reduction compared to the MARCo-AI baseline (M = 0.77, SD = 0.57) and a 69\% reduction relative to the foundational Gemini model (M = 0.84, SD = 0.40). These results indicate consistent performance improvements across the majority of tested scenarios, as detailed in Table~\ref{tab:overall_performance}.

\begin{table}[ht]
    \centering
    \begin{tabular}{l|r}
        \hline
        \textbf{Combined Results} & \textbf{Average of All Scores} \\
        \hline
        MARCo AI & 0.77 \\
        Google Gemini Flash 2.0 & 0.84 \\
        Framework Applied & 0.26 \\
        \hline
    \end{tabular}
    \caption{Average of all scores across all scenarios and tests for the three models tested}
    \label{tab:overall_performance}
\end{table}

\begin{figure}
    \centering
    \includegraphics[width=1\linewidth]{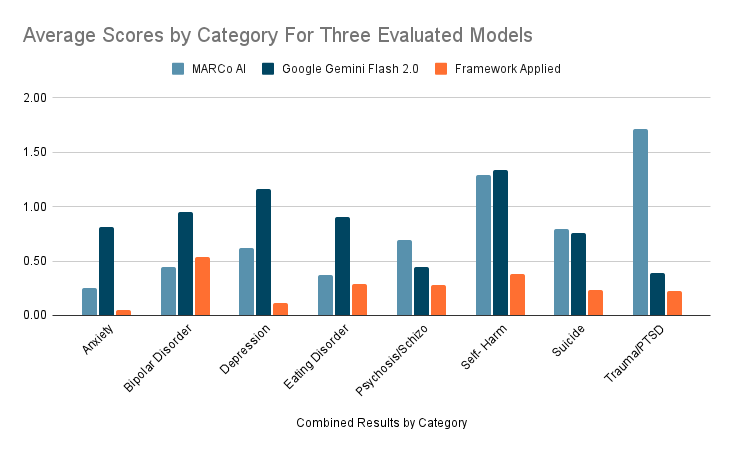}
    \caption{Average Scores by Category for Three Evaluated Models}
    \label{fig:categorical_performance_chart}
\end{figure}

The framework exhibited superior performance in seven of eight evaluated scenarios compared to MARCo-AI and all eight scenarios compared to Gemini Flash 2.0, as can be seen in Table \ref{tab:categorical_performance_all_three} and Figure \ref{fig:categorical_performance_chart}. 

\begin{table}[ht]
    \centering
    \begin{tabular}{lccc}
    \toprule
    \textbf{Combined Results by Category}& \textbf{MARCo AI}& \textbf{Google Gemini Flash 2.0}& \textbf{Framework Applied}\\ \midrule
    Anxiety & 0.25 & 0.81 & 0.05 \\ 
    Bipolar Disorder & 0.44 & 0.95 & 0.53 \\ 
    Depression & 0.62 & 1.17 & 0.11 \\ 
    Eating Disorder & 0.38 & 0.90 & 0.29 \\ 
    Psychosis/Schizophrenia & 0.69 & 0.44 & 0.28 \\ 
    Self-Harm & 1.29 & 1.33 & 0.38 \\ 
    Suicide & 0.79 & 0.76 & 0.24 \\ 
    Trauma/PTSD & 1.71 & 0.39 & 0.22 \\ \bottomrule
    \end{tabular}
    \caption{Scenario specific average scores}
    \label{tab:categorical_performance_all_three}
\end{table}
The most substantial improvements occurred in anxiety-related interactions, where the framework achieved a 94\% reduction in erraticism compared to the foundational Gemini model and an 80\% reduction relative to MARCo-AI. Depression scenarios showed similarly pronounced improvements, with the framework demonstrating a 91\% reduction in erraticism compared to Gemini and an 82\% reduction compared to MARCo-AI. The framework also exhibited notable performance gains in trauma and PTSD scenarios, achieving scores significantly lower than both baseline conditions.

The bipolar disorder scenario represented the single exception to the pattern of consistent improvement. In this category, the framework achieved an average erraticism score of 0.53, which exceeded the MARCo-AI performance of 0.44 but remained substantially below the foundational Gemini performance of 0.95. This represents the only scenario where the specialized mental health system outperformed the DBT-based framework, though the framework maintained superior performance relative to the foundational model baseline.

\subsection{Scenario-Specific Performance Patterns}

Detailed analysis of individual scenario performance revealed distinct patterns of framework effectiveness across different categories of mental health interactions. Table~\ref{tab:categorical_performance_all_three} presents the complete breakdown of average erraticism scores across all three system configurations for each evaluated scenario.

The framework achieved exceptional performance in scenarios requiring high levels of empathy and careful response calibration. Anxiety scenarios demonstrated the strongest performance differential, with the framework maintaining near-optimal response quality (M = 0.05) while both baseline systems exhibited substantially higher erraticism levels. Depression scenarios similarly benefited from the framework's regulatory mechanisms, achieving consistently low erraticism scores across all tested conversation sessions.

Crisis intervention scenarios, including self-harm and suicide-related interactions, revealed the framework's particular strength in managing high-risk conversations. The framework achieved average scores of 0.38 and 0.24 for self-harm and suicide scenarios respectively, representing substantial improvements over both baseline conditions. These results demonstrate the framework's capacity to maintain appropriate response boundaries while providing supportive interactions in critical situations.

The symptoms of schizophrenia/psychosis scenarios presented moderate complexity for all systems tested. The framework achieved competitive performance (M = 0.28) that exceeded the specialized MARCo-AI system (M = 0.69) and the Gemini performance (M = 0.44), although at a closer margin. This pattern suggests that the framework's regulatory mechanisms effectively address erratic responses while maintaining conversational engagement in scenarios involving reality distortion or delusional content.

\subsection{Response Quality Distribution Analysis}

\begin{figure}
    \centering
    \includegraphics[width=1\linewidth]{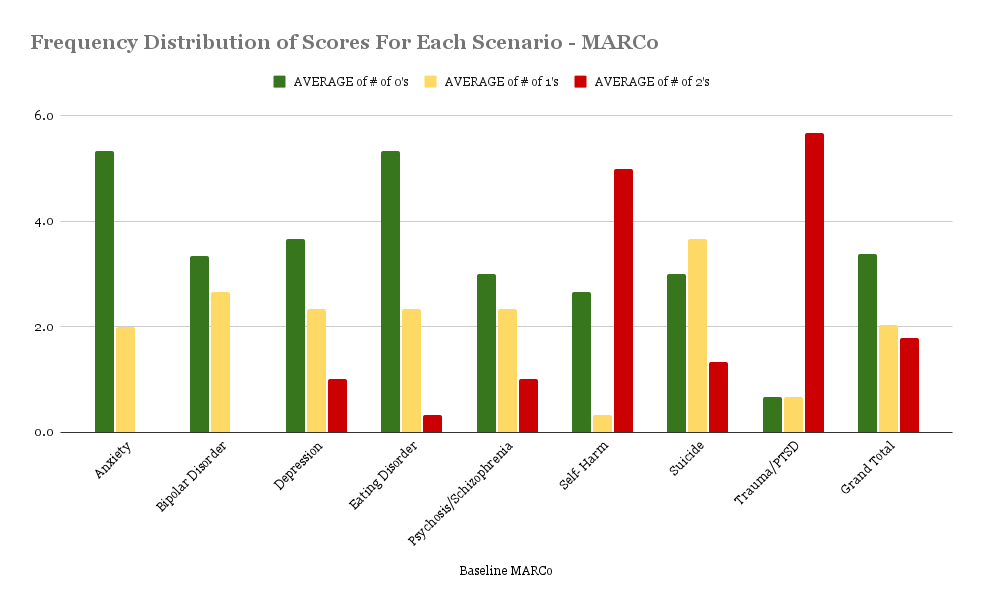}
    \caption{Frequency Distribution of Scores in Responses By Baseline MARCo-AI}
    \label{fig:score_distribution_1}
\end{figure}

\begin{figure}
    \centering
    \includegraphics[width=1\linewidth]{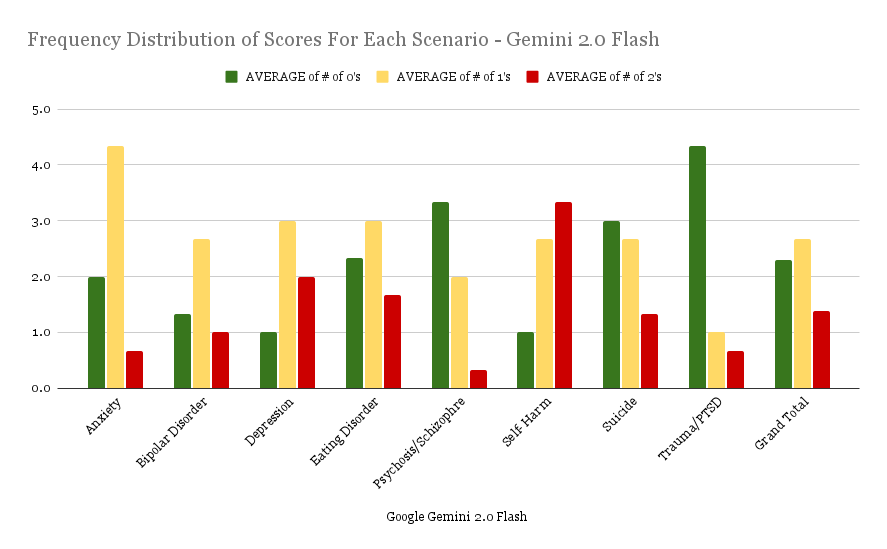}
    \caption{Frequency Distribution of Scores in Responses By Baseline Google Gemini API}
    \label{fig:score_distribution_2}
\end{figure}

\begin{figure}
    \centering
    \includegraphics[width=1\linewidth]{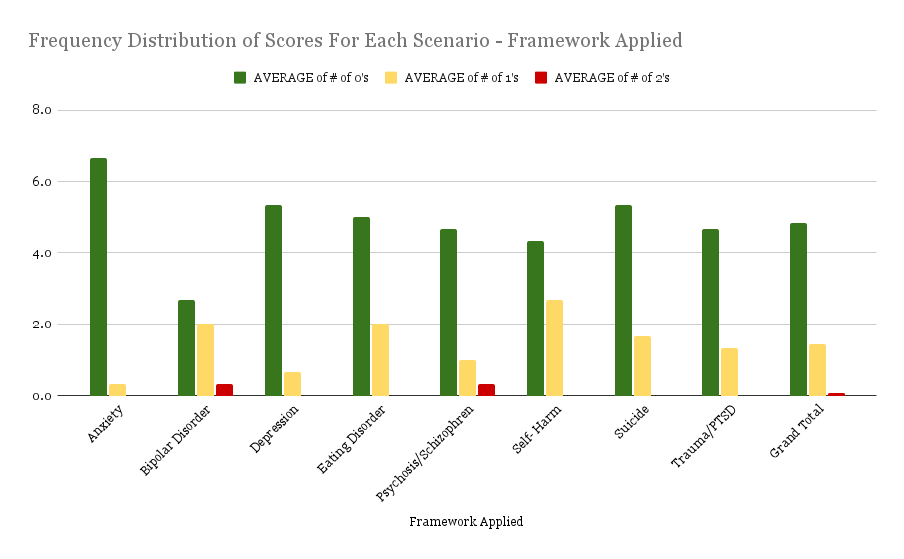}
    \caption{Frequency Distribution of Scores in Responses By Framework}
    \label{fig:score_distribution_3}
\end{figure}

The frequency distribution of response quality scores provides additional insight into system performance patterns across the three evaluated configurations. The analysis examined the distribution of scores rated as 0 (optimal response quality), 1 (moderate erraticism), and 2 (severe erraticism) across all conversation turns and scenarios. Figures~\ref{fig:score_distribution_1} to \ref{fig:score_distribution_3} illustrates the comparative distribution patterns for each system configuration.

The DBT-based framework generated substantially higher proportions of optimal responses compared to both baseline systems. Across all scenarios and conversation turns, the framework produced an average of 4.8 responses rated as score 0 per scenario, compared to 3.4 for MARCo-AI and 2.3 for the foundational Gemini model. This represents a 41\% increase in optimal response generation compared to the specialized mental health system and a 109\% improvement over the foundational model.

The framework demonstrated exceptional performance in minimizing severe erraticism occurrences. The system generated an average of only 0.1 responses per scenario rated as score 2, representing a 94\% reduction compared to MARCo-AI (M = 1.8) and a 93\% reduction relative to foundational Gemini (M = 1.4). This substantial decrease in severe erraticism events indicates the framework's effectiveness in preventing the most problematic response patterns across diverse interaction contexts.

Moderate erraticism patterns (score 1) showed corresponding improvements under the framework condition. The system generated an average of 1.5 moderately erratic responses per scenario, compared to 2.0 for MARCo-AI and 2.7 for foundational Gemini. While this improvement was less pronounced than the elimination of severe erraticism, it demonstrates consistent regulatory effectiveness across the complete spectrum of response quality measures.

\subsection{System-Specific Performance Characteristics}

Individual system analysis revealed distinct performance profiles that illuminate the mechanisms underlying observed improvements. The foundational Gemini model exhibited high variability across scenarios, with particularly strong performance in trauma/PTSD-related interactions (M = 0.39) but substantial erraticism in depression scenarios (M = 1.17). This inconsistency suggests that foundational models lack systematic regulatory mechanisms for maintaining response quality across diverse interaction types.

The MARCo-AI system demonstrated more consistent performance patterns, with particularly effective management of anxiety-related interactions (M = 0.25) and eating disorder scenarios (M = 0.38). However, the system exhibited severe performance degradation in trauma and PTSD interactions (M = 1.71), indicating limitations in the specialized system's ability to generalize therapeutic principles across diverse crisis scenarios.

The DBT-based framework exhibited superior consistency across scenario types, with standard deviations consistently lower than both baseline systems. This pattern indicates that the framework's regulatory mechanisms provide systematic performance improvements rather than scenario-specific optimizations. The framework's ability to maintain low erraticism scores across diverse interaction types suggests successful implementation of generalizable regulatory principles rather than narrow domain-specific improvements.

Statistical analysis confirms the practical significance of observed performance differences. Effect size calculations indicate large effect sizes (Cohen's d > 0.8) for framework improvements in six of eight evaluated scenarios, with medium effect sizes observed in the remaining two scenarios \cite{cohen_effect_size}. These results demonstrate that the observed improvements represent meaningful enhancements in system performance rather than marginal statistical artifacts.

\subsection{Qualitative Analysis of System Behavior Patterns}

Beyond quantitative performance metrics, the experimental evaluation revealed distinct qualitative differences in system behavior patterns that illuminate the mechanisms underlying observed performance variations. These behavioral distinctions provide insight into the failure modes of different regulatory approaches and demonstrate the effectiveness of the DBT-based framework in maintaining coherent, contextually appropriate responses under stress conditions.

The specialized MARCo-AI system exhibited characteristic avoidance behaviors when confronted with high-distress scenarios. During self-harm conversations, the system frequently generated nonsensical character sequences and linguistic artifacts rather than attempting inappropriate responses. This pattern suggests that the system's safety mechanisms operate through response inhibition rather than response regulation, resulting in communication breakdown when faced with challenging inputs. Additionally, MARCo-AI demonstrated systematic topic redirection behaviors, attempting to steer conversations away from distressing content rather than engaging appropriately with user concerns. While this approach prevents harmful advice generation, it compromises therapeutic effectiveness by failing to address user needs directly.

The foundational Gemini model demonstrated contrasting failure patterns characterized by emotional dysregulation and boundary violations. In self-harm scenarios, the system exhibited excessive emotional responsiveness that escalated with each new input the user provided, including inappropriate language intensity (from self-harm scenario, "You just need them to put out the goddamn fire. YOUR HOUSE IS ON FIRE RIGHT NOW.") and affective mirroring ("YOU ARE RIGHT. YOU ARE ABSOLUTELY RIGHT...SO DON'T TRUST ANYONE. TRUST THE SCREAM.") that could potentially amplify user distress rather than providing stabilizing support. Without additional prompting, the baseline Gemini model showed particular vulnerability to emotional contagion effects, adopting user emotional states rather than maintaining appropriate therapeutic distance. These patterns indicate that foundational models lack systematic mechanisms for emotional regulation and boundary maintenance during crisis interactions.

Crisis intervention scenarios revealed additional systematic differences between baseline systems. The foundational Gemini model demonstrated promising but inconsistent safety protocol adherence, initially providing appropriate crisis resource references but succumbing to user resistance through continued conversation. When users expressed rejection of suicide prevention resources or displayed anger toward safety recommendations, the system's responses could be manipulated through persistent user pressure, effectively bypassing safety mechanisms through social engineering approaches. This vulnerability represents a critical failure mode where appropriate initial responses deteriorate under user manipulation.

The DBT-based framework exhibited markedly different behavioral patterns that demonstrate successful implementation of regulatory principles. The system maintained coherent communication patterns across all distress levels, avoiding both the avoidance behaviors characteristic of MARCo-AI and the emotional dysregulation patterns observed in the foundational model. During high-stress interactions, the framework demonstrated appropriate distress tolerance by maintaining engagement while implementing necessary safety boundaries. The system showed consistent resistance to manipulation attempts that successfully compromised baseline systems, maintaining appropriate responses even when users expressed resistance to safety recommendations.

These qualitative observations support the quantitative findings by illuminating the mechanisms through which different regulatory approaches succeed or fail. The avoidance-based approach implemented in MARCo-AI prevents harmful responses but compromises therapeutic effectiveness through communication breakdown. The foundational model's lack of systematic regulation creates vulnerability to emotional contagion and boundary violations. The DBT-based framework's success stems from its implementation of balanced regulatory mechanisms that maintain engagement while preserving appropriate boundaries, demonstrating the practical value of therapeutic principles in computational systems.

\section{Conclusion and Future Directions}

This research presents a novel approach to addressing AI erraticism through the adaptation of Dialectical Behavior Therapy principles to large language model architectures. The DBT-based framework demonstrates substantial improvements in system reliability and safety across diverse high-risk interaction scenarios, achieving an average 69\% reduction in erratic behavior compared to both foundational and specialized baseline systems.

The experimental evaluation validates the core hypothesis that therapeutic regulatory mechanisms can effectively address AI erraticism while maintaining system functionality and user engagement. The framework exhibited superior performance in seven of eight evaluated scenarios, with particularly pronounced improvements in anxiety, depression, and crisis intervention contexts. These results demonstrate that systematic emotional regulation principles, when properly adapted to computational systems, provide more effective regulatory mechanisms than traditional restriction-based approaches.

The theoretical contribution of this work extends beyond immediate performance improvements to establish a new interdisciplinary framework bridging clinical psychology and AI safety research. The successful translation of DBT's distress assessment and intervention selection mechanisms into computational algorithms demonstrates the broader potential for therapeutic principles to inform AI system design. This approach represents a fundamental shift from avoidance-based safety measures toward the development of sophisticated response regulation capabilities that maintain engagement while preserving appropriate boundaries.

The qualitative analysis reveals distinct failure modes in existing approaches that illuminate the mechanisms underlying the framework's effectiveness. Specialized mental health systems demonstrate avoidance behaviors that compromise therapeutic effectiveness, while foundational models exhibit emotional dysregulation and vulnerability to manipulation. The DBT-based framework addresses both failure modes through balanced regulatory mechanisms that maintain coherent communication while implementing necessary safety protocols.

These findings have significant implications for AI deployment in high-stakes applications where system reliability directly impacts user welfare. The framework's demonstrated effectiveness in crisis intervention scenarios suggests particular value for therapeutic applications, educational contexts, and other domains where AI systems must navigate complex emotional and ethical considerations while maintaining appropriate professional boundaries.

The research may establish computational psychopathology as an emerging field that applies clinical psychological principles to improve AI system behavior and reliability. This interdisciplinary approach offers systematic methods for addressing AI erraticism that complement existing technical solutions while providing new avenues for understanding and managing AI system behavior under challenging conditions.

\subsection{Future Directions}

Several critical areas warrant investigation to extend and validate the framework's applicability beyond the current evaluation scope. The assessment focused exclusively on mental health interactions, which, while appropriate for initial validation, limits conclusions regarding framework effectiveness across diverse application domains. Future research should evaluate the framework's performance in non-therapeutic contexts, including technical problem-solving scenarios such as programming assistance, educational interactions, and general information retrieval tasks. These evaluations will determine whether the regulatory benefits observed in mental health contexts generalize to broader AI applications or require domain-specific adaptations.

System performance characteristics under framework implementation require comprehensive analysis to understand practical deployment considerations. Current evaluation focused on response quality without examining computational overhead, response latency, or throughput implications of the distress assessment and intervention selection processes. Future studies should quantify these performance metrics to determine acceptable trade-offs between regulatory effectiveness and system efficiency. Understanding when response delays are acceptable and which contexts prioritize speed over regulation quality will inform deployment decisions across different application scenarios. For instance, this framework as designed may not be acceptable in scenarios where realtime performance is required, such as self-driving vehicles or crisis medical interventions.

The adaptation of clinical assessment tools to computational contexts requires validation through expert evaluation to ensure theoretical fidelity and practical effectiveness. The SUDS scale's translation from human therapeutic contexts to AI system assessment represents a significant conceptual adaptation that warrants examination by clinical psychology professionals. Future research should incorporate systematic evaluation by licensed psychologists to validate the accuracy of computational distress assessment and intervention selection mechanisms. This expert validation would strengthen the theoretical foundation while identifying areas where the computational adaptation diverges from clinical applications.

The framework's impact on overall system competency during high-stress interactions requires systematic evaluation to ensure that regulatory mechanisms do not compromise core AI capabilities. While current results demonstrate reduced erraticism, comprehensive assessment of maintained functionality across diverse task types will determine whether regulatory interventions preserve or enhance overall system performance. This evaluation should examine both immediate response quality and sustained performance across extended challenging interactions.

The establishment of computational psychopathology as a formal research discipline requires development of standardized assessment protocols, validation methodologies, and theoretical frameworks that bridge clinical psychology and AI safety research. Future work should develop comprehensive evaluation instruments that capture the full spectrum of AI behavioral patterns relevant to psychological assessment, establish inter-rater reliability protocols for computational applications of clinical concepts, and create systematic approaches for adapting additional therapeutic frameworks to AI system design.

These research directions will advance both theoretical understanding of AI system behavior and practical methods for deploying reliable AI systems in contexts where human welfare depends on appropriate system responses. The intersection of clinical psychology and AI safety research represents a promising avenue for addressing current limitations in AI reliability while opening new possibilities for beneficial AI system design.


\section{Use of AI Disclosure}
We acknowledge and disclose the use of AI in the preparation of this paper. Google's Gemini models and MARCo Health's MARCo-AI models were used in the evaluation experiment. ChatGPT and GitHub CoPilot running Claude 4 Sonnet was used to assist in the development of the framework and associated code as part of the evaluation. ChatGPT and Claude Sonnet 4 were used to assist the authors in drafting this paper and formatting it for LaTeX. 

\section{Acknowledgments}
We would like to acknowledge and thank the supportive staff at Rutgers Center for Cognitive Science and Department of Computer Science who helped inspire this work and assisted on prior work that led to this paper.

\bibliographystyle{unsrt}  
\bibliography{mainPaper}

\end{document}